\documentclass[fleqn, usenatbib]{mnras}

\usepackage[T1]{fontenc}
\usepackage{ae,aecompl}

\usepackage{graphicx}	
\usepackage{amsmath}	
\usepackage{amssymb}	
\usepackage{xcolor}

\usepackage{etoolbox}
\makeatletter
\newcount\c@additionalboxlevel
\newcount\c@maxboxlevel
\patchcmd\@combinedblfloats{\box\@outputbox}{%
  \stepcounter{additionalboxlevel}%
  \box\@outputbox
}{}{\errmessage{\noexpand\@combinedblfloats could not be patched}}

\AtBeginShipout{%
  \ifnum\value{additionalboxlevel}>\value{maxboxlevel}%
    \typeout{Warning: maxboxlevel might be too small, increase to %
      \the\value{additionalboxlevel}%
    }%
  \fi 
  \@whilenum\value{additionalboxlevel}<\value{maxboxlevel}\do{%
    \typeout{* Additional boxing of page `\thepage'}%
    \setbox\AtBeginShipoutBox=\hbox{\copy\AtBeginShipoutBox}%
    \stepcounter{additionalboxlevel}%
  }%
  \setcounter{additionalboxlevel}{0}%
}
\makeatother

\def\be{\begin{equation}}
\def\ee{\end{equation}}

\newcommand\quotes[1]{``{#1}"}

\def\zsun{{\rm Z}_{\odot}}

\definecolor{apcolor}{HTML}{b3003b}
\definecolor{afcolor}{HTML}{800080}
\definecolor{lvcolor}{HTML}{DF7401}
\title[XDR in high-$z$ galaxies]{Impact of X-rays on CO emission from high-$z$ galaxies}  

\author[Vallini et al.]
{L. Vallini$^{1}$\thanks{E-mail: vallini@strw.leidenuniv.nl (LV)},
A. G. G. M. Tielens$^{1}$,
A. Pallottini$^{2,3}$,
S. Gallerani$^{3}$,
C. Gruppioni$^{4}$,
\newauthor
S. Carniani$^{3}$
F. Pozzi$^{5}$,
M. Talia$^{5}$
\\
$^{1}$Leiden Observatory, Leiden University, PO Box 9500, 2300 RA Leiden, The Netherlands.\\
$^{2}$Centro Fermi, Museo Storico della Fisica e Centro Studi e Ricerche \quotes{Enrico Fermi}, Piazza del Viminale 1, Roma, I-00184, Italy\\
$^{3}$Scuola Normale Superiore, Piazza dei Cavalieri 7, I-56126 Pisa, Italy\\
$^{4}$INAF- Osservatorio Astronomico di Bologna, Via Gobetti 93/3, I-40129, Bologna, Italy\\
$^{5}$Dipartimento di Fisica e Astronomia, Universit`a di Bologna, Via Gobetti 93/2, I-40129, Bologna, Italy.}

\date{Accepted XXX. Received YYY; in original form ZZZ}

\pubyear{2018}
\begin{document}
\label{firstpage}
\pagerange{\pageref{firstpage}--\pageref{lastpage}}
\maketitle
\begin{abstract}
We study the impact of active galactic nuclei (AGN) on the CO Spectral Line Energy Distribution (SLED) of high-$z$ galaxies. In particular, we want to assess if the CO SLED can be used as a probe of AGN activity.
To this purpose, we develop a semi-analytical model that takes into account the radiative transfer and the clumpy structure of giant molecular clouds where the CO lines are excited, their distribution in the galaxy disk, and the torus obscuration of the AGN radiation. We study the joint effect on the CO SLED excitation of (i) the X-ray luminosity from the AGN ($L_{X}$), (ii) the size of the molecular disk, (iii) the inclination angle ($\Omega$) of the torus with respect to the molecular disk, and (iv) the GMC mean density. We also discuss the possibility of an enhanced Cosmic Ray Ionization Rate (CRIR). We find that the X-ray Dominated Region (XDR) generated by the AGN in every case enhances the CO SLED for $J>5$, with increasing excitation of high-$J$ CO lines for increasing X-ray luminosity. 
Because high-$z$ galaxies are compact, the XDR region typically encloses the whole disk, thus its effect can be more important with respect to lower redshift objects. The impact of the XDR can be disentangled from an enhanced CRIR either if $L_X>10^{44} \rm \,erg\, s^{-1}$, or if $\Omega \geq 60^{\circ}$. We finally provide predictions on the CO(7--6)/[CII] and CO(17--16)/[CII] ratios as a function of $L_X$, which can be relevant for ALMA follow up of galaxies and quasars previously detected in [CII].
\end{abstract}




\section{Introduction}
The advent of ALMA has opened a window on the characterization of the cold gas in galaxies during the Epoch of Reionization \citep[EoR,][]{carilli2013}. Being sensitive at sub-millimeter wavelengths, ALMA allows the detection of the (redshifted) far-infrared emission of metal cooling lines (e.g. [CII] at 158 $\rm \mu m$), and carbon monoxide (CO) rotational lines ($J>5$, from $z>6$) tracing the neutral and molecular gas, respectively.
While the [CII] 158 $\mu m$ transition -- being the most luminous line in the FIR band \citep{stacey1991} -- has started to be routinely detected in normal star forming galaxies ($\mathrm{ SFR\leq 100\, M_{\sun} yr^{-1}}$) at the end ($z\approx 6-7$) of the EoR \citep[e.g.][]{capak2015, willott2015, maiolino2015, pentericci2016, jones2017, matthee2017, smit2018, carniani2018, matthee2019}, observations of CO rotational lines in normal galaxies are still rare \citep[][]{dodorico2018, pavesi2018b}. Almost all CO line detection from $z>6$ are reported in luminous quasars \citep[e.g.][]{bertoldi2003, walter2012, wang2010, wang2011, stefan2015, gallerani2014, venemans2017, feruglio2018, novak2019, wang2019,carniani2019:CO}, and dusty star-forming (SFR$>1000 \rm M_{\odot}\, yr^{-1}$) galaxies \citep[e.g.][]{strandet2017, marrone2018, pavesi2018a}.
However, despite the difficulties in the detection of CO lines from normal galaxies, and considering that other molecular gas tracers such as $\rm H_2$ rotational lines will become observable from high-$z$ only with the advent of SPICA \citep{spinoglio2017, egami2018}, the CO Spectral Line Energy Distribution (CO SLED) -- i.e. the CO line luminosity as a function of rotational quantum number $J$ -- remains a valuable probe of the conditions of molecular gas in such primeval objects \citep{vallini2018}.

The relative luminosity of low-$J$ ($J\leq3$), mid-$J$ ($4\leq J\leq 7$) and high-$J$ ($J\geq 8$) transitions is a proxy of the relative abundance of cold low-density molecular gas ($T \approx 20-50$ K, $n=100-1000 \, \rm cm^{-3}$), traced by low-$J$ CO rotational transitions, versus that of warm high-density molecular gas ($T=150-1000\, \rm K$, $n>10^4 \, \rm cm^{-3}$), in which increasingly high-$J$ lines are excited \citep[e.g.][]{obreschkow2009, lagos2012,mashian2015, rosenberg2015}. The high-$J$ CO lines are even more excited in extreme environments such e.g. in the vicinity of active galactic nuclei (AGN), or in shock heated molecular gas \citep{meijerink2007, schleicher2010, gallerani2014, rosenberg2015, mashian2015, popping2016, pozzi2017, mingozzi2018, talia2018}.

The influence of the central AGN on the CO excitation is generally limited to the inner $\approx 0.5\, \rm{kpc}$ of the galaxy \citep[e.g.][]{vanderwerf2010, pozzi2017, mingozzi2018}. In this region, the GMCs are exposed both to the intense radiation in the far-ultraviolet (FUV, $6.0 < h\nu < 13.6$ eV) produced by the star formation -- turning cloud surfaces into Photodissociation Regions \citep[PDRs,][]{hollenbach1999} -- and to X-ray photons ($h \nu > 1$ keV) from the AGN. X-ray photons penetrate deeper into the GMCs with respect to FUV, creating the so called X-ray Dominated Regions \citep[XDRs,][]{maloney1996}. The thermal and chemical structures of XDRs and PDRs are different because in XDRs the molecular gas is kept warmer up to larger (column) densities \citep{maloney1996}. Hence, while the CO SLED in PDRs peaks around CO(5--4) and decreases rapidly for higher J transitions, in XDRs, the CO emission continues to rise to CO(10--9) or even higher transitions.
\citet{schleicher2010} pointed out that even with the unprecedented spatial resolution of ALMA (up to $\approx 300-500 \rm \,pc$ at $z\approx 6$), the disentanglement and assessment of the AGN influence on the molecular gas excitation might be hampered by the difficulty in detecting in a reasonable amount of time the CO emission from the central region only. The key point, however, is that galaxies at high-$z$ are very compact and thus, unlike low-$z$ Seyfert galaxies, the global CO SLED might carry the imprint of the XDR component.
For example, observational studies based on photometrically selected galaxies at $z>6$ have shown that most of them are barely spatially resolved and with typical rest-frame UV sizes of $<1$ kpc \citep[e.g.][]{stanway2004, oesch2010, cowie2011, jiang2013, shibuya2019}. 
Moreover, recent state-of-the-art cosmological zoom-in simulations show that we expect all the molecular gas residing in compact disks with typical radii of $r\approx 0.5$ kpc \citep{pallottini2017b, vallini2018}, i.e. comparable with that traced by UV light.
Therefore the fraction of GMCs residing in the central $\approx 500\, \rm pc$ region of a typical high-$z$ galaxy is higher than that of typical local Seyfert galaxies which have molecular disks of $r\approx 2-2.5 \, \rm kpc$ \citep{downes1998, casasola2017}. 
This means that it might be easier to catch the imprint of an XDR on the global CO SLED for a high-$z$ galaxy rather 
than for a local Seyfert galaxy. In the latter case in fact the XDR contribution to the CO excitation is smeared out if the global CO SLED, instead of the spatially resolved one, is observed \citep{vanderwerf2010, mashian2015, mingozzi2018}.

The goal of this paper is to assess the effect of the central AGN on the global CO SLED to identify which J transitions offer the best probes of irradiation of molecular clouds by X-rays produced by accretion disk surrounding the central black hole.
Although the method developed here might be applied to any galaxy, here we will ultimately focus our discussion on $z>6$ sources.
Note that the detection, even indirect, of low-luminosity AGN is potentially extremely relevant to solve the puzzling issue of formation of supermassive black holes at $z>6$ that strongly challenges all the standard theories of black hole growth \citep[e.g.][]{volonteri2010, valiante2017, gallerani2017rev}. In this paper, we extend the CO excitation model for high-$z$ galaxies presented in \citet{vallini2018} that considers only the FUV radiation as possible excitation source, by including the presence of X-rays illuminating the GMCs.

The paper is organised as follows: in Sec. \ref{sec:Model_description} we present the analytical model developed to compute the CO SLED. In Sec. \ref{sec:lowzvalidation} we validate the model in a low-$z$ well studied Seyfert galaxy. In Sec.s \ref{sec:parameter_effect} and \ref{sec:cii_co_QSO} we adopt it to provide predictions on high-$z$ galaxies. Finally, in Section \ref{sec:conclusions}, we discuss the caveats of our model and draw our conclusions.

\begin{figure*}
\centering
\includegraphics[scale=0.6]{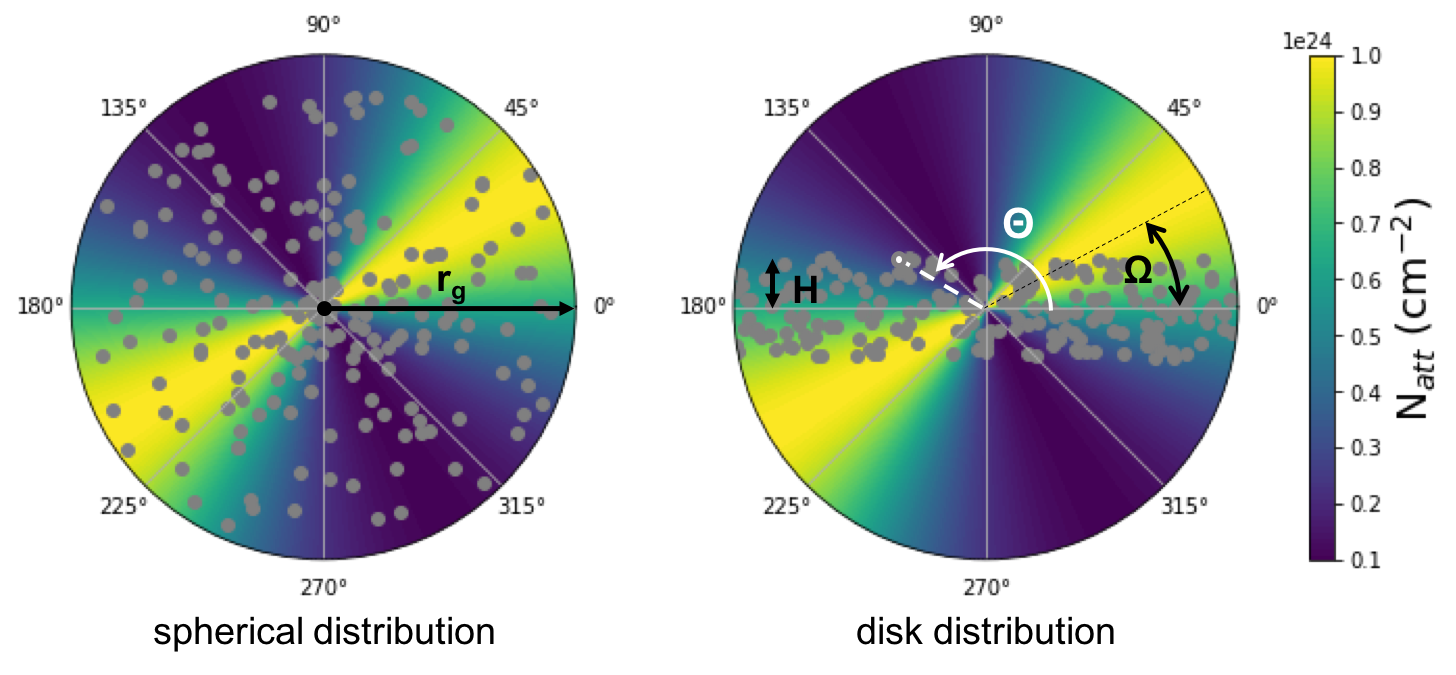}
\caption{Sketch of the GMC distribution (grey dots) assumed through this work to model the emission from molecular gas. The plot shows an edge-on view of the model. The free parameters in the modelling are: i) the maximum galactocentric radius ($r_g$) up to which the GMCs are distributed, ii) the height of the disk ($H$), and iii) the tilt of the obscuring torus ($\Omega$, in this specific case $\Omega=30^{\circ}$) with respect to the galactic plane. The obscuration produced by the torus in the direction ($\Theta$) of each GMC is color coded and parametrised according to eq. \ref{eq:obscuration}. \label{fig:sketch_model}}
\end{figure*}

\section{Model description}\label{sec:Model_description}

To elucidate the impact of AGN activity on the CO excitation we develop a model in which the mo\-le\-cular gas in a galaxy is distributed in GMCs illuminated by the radiation coming both from the central AGN, and by that provided by star formation, as depicted in the skecth in Figure \ref{fig:sketch_model}.
Properly trea\-ting the internal density field of molecular clouds is pivotal when computing the emission of high-$J$ CO rotational lines because they have high critical densities, and thus they trace the densest regions of GMCs. Similary to \cite{vallini2017, vallini2018} we take into account of this aspect, by considering the probability distribution function of GMC densities as controlled by the Mach number ($\mathcal{M}$) of the turbulence \citep{mckee2007, federrath2013, padoan2014}.
The CO emission from the UV illuminated surface of the molecular clouds will be computed adopting a full PDR calculation (see Section \ref{sec:FUVillumination} and \ref{sec:PDR_modelling}), while the emission from X-ray illuminated GMCs is computed through XDR models (see Sec. \ref{sec:Xrayillumination} and \ref{sec:XDR_modelling}). We will also test the effect of an increased cosmic ray flux (Sec. \ref{sec:RT_CRs}) on the CO emission predicted by our model.

\subsection{The molecular cloud distribution}\label{sec:galaxy_distribution}

We distribute the GMCs around the central X-ray source (with intrinsic X-ray luminosity, $L_{\rm X}$), considering two cases: a spherical distribution -- see Fig. \ref{fig:sketch_model}, left panel -- or a disk-like distribution, see Fig. \ref{fig:sketch_model}, right panel. The spherical distribution is meant to represent the case of irregular morphology which is often observed in high-$z$ sources \citep{jiang2013}. For the spherical case, we distribute the GMCs uniformly over the volume.

The surface density profile for the molecular disk case is instead assumed to follow an exponential relation of the form:
\begin{equation}
\Sigma (r) = \Sigma _0 \, {\rm exp} \left( -\frac{r}{r_g}\right),
\end{equation}
where we assume $\Sigma_0 \approx 200 \,\rm M_{\odot}\, pc^{-2}$ as fiducial, in agreement with the characteristic disk surface density found in $z\approx 6$ galaxies \citep{pallottini2017b, pallottini2019}. The other parameters of our model are the maximum galactocentric radius up to which the clouds are distributed ($r_g$), and -- for the disk only -- the height of the disk, $H$ (see Fig. \ref{fig:sketch_model}, right).

\subsection{FUV illumination of the GMCs}\label{sec:FUVillumination}

In our model the GMCs are illuminated by the radiation provided by the star formation. The stellar radiation is parametrised in term of the interstellar FUV flux, $G_0$, normalised to that in the solar neighbourhood ($G_{0,MW}\approx 1.6\times 10^{-3}\rm {erg\, cm^{-2}\, s^{-1}}$, \citealt[][]{habing1968}). A constant value $G_0$ is assumed to illuminate all the GMCs and it scales linearly with the putative SFR of the galaxy:
\begin{equation}
    G_0 = \left( \frac{\mathrm SFR}{\mathrm SFR_{MW}}\right) G_{0, MW}.
    \label{eq:g0}
\end{equation}
where we set $SFR_{MW}= 1\, \rm M_{\odot} \, yr^{-1}$, consistently with \citet{habing1968} for the calculation of the mean FUV field in the solar neighborhood. Note that such approximation of uniform flux is in broad agreement with the results of radiative transfer calculation of high-redshift galaxy simulations \citep[e.g.][]{behrens2018, pallottini2019}.

\subsection{X-ray illumination of the GMCs}\label{sec:Xrayillumination}
If an AGN is present then its radiation has an effect on the molecular gas heating via X-rays.
In the so called ``unified model'' the accretion disk of the AGN, responsible for the production of X-ray photons, is surrounded by a toroidal distribution of dust and gas that can be optically thick \citep[][for a recent review]{hickox2018}.
Due to the anisotropic nature of the torus, the obscuration is function of the polar angle of GMCs ($\Theta$): along some lines of sights (l.o.s.), the accretion-disk emission is directly detected because the photons propagate freely, while for other l.o.s. the emission is obscured by the torus.
This impact the X-ray flux impinging the GMC surfaces.
The torus can be tilted with respect to the galaxy disk (located at $\Theta=0$, see Fig. \ref{fig:sketch_model}) by an angle $\Omega$.

The model AGN host galaxy sketched in Fig. \ref{fig:sketch_model} consists of the molecular gas, distributed in GMCs either spherically or along a disk, plus an axisymmetrical X-ray absorbing torus surrounding the central BH. 
The molecular gas column density along each line of sight depends on the actual distribution (and shadowing) of the GMCs. We approximate it as follows:
\begin{equation}
\label{eq:absorption_gas}
    N_{gas}(r') \approx N_{\rm H} \int_0^{r'} dN_{GMC}(r) dr
\end{equation}
where $N_{\rm H}$ is the mean column density of each GMC, and $\int_0^{r'} dN_{GMC}(r) dr$ is the mean number of GMCs from the center.
This is negligible only if the filling factor of GMCs is small, as typical column density of a GMC is in the range $N_H=1-2 \times 10^{22} \rm {cm^{-2}}$, i.e. much smaller than that provided by the torus.

For what concerns the torus, following the model developed by \citet{galliano2003} to explain the CO emission from the inner ($r<500\, \rm pc$) region of the nearby AGN NGC 1068, the attenuating column density ($N_{\rm att}$) depends on the angle $\Theta$ as:
\begin{equation}
N_{\rm att}(\Theta) = N_{\rm att}(0) 10^{cos^{1.5} (\Theta - \Omega)},
\label{eq:obscuration}
\end{equation}
and it is depicted as a color gradient in Fig. \ref{fig:sketch_model}. In eq. \ref{eq:obscuration}, $N_{\rm att}(0)$ is the equatorial thickness of the X-ray absorber and $0^{\circ}\leq \Omega \leq 90^{\circ}$ is the tilt angle between the plane of the torus and that of galaxy. 
Assuming an intrinsic X-ray luminosity $L_{X}$ in the energy range between 1 and 100 keV, then the attenuated X-ray flux at the surface of a GMC located at a distance $r$ from the central BH, can be written as \citep[see][for the complete derivation]{maloney1996}:
\begin{equation}
F_{X}= \frac{L_X}{4 \pi r^2 N_{22}^{0.9}},
\label{eq:fx}
\end{equation}
where $N_{22}=(N_{\rm att}+N_{gas})/10^{22}~ \rm cm^{-2}$. GMCs in the galactic disk, at a distance $r$ from the central BH, are exposed to increasing X-ray impinging fluxes at fixed $L_{X}$, for $\Omega$ growing from 0$^\circ$ to 90$^\circ$.

\subsection{The internal structure of GMCs}\label{sec:internal_structure}

All clouds considered in our model are identical to the fiducial GMCs described in \citet{vallini2018} which developed a semi-analytical model that -- given the internal density field of the GMC and the impinging radiation at its surface -- computes the CO luminosity of the various J transitions ($L_{\rm CO, J}$).
The fiducial model clouds are assumed to be spherical, with a radius $r_{\rm GMC}=15 \rm \, pc$, total mass $M_{\rm GMC}\approx 1 \times 10^{5} \rm \, M_{\odot}$, and mean density $n_0\approx 300 \rm cm^{-3}$. Turbulence and self gravity are also accounted for, because they impact substantially the internal density field of the GMCs \citep{larson1981, mckee2007}. More precisely, the turbulence produces an internal density field characterized by a lognormal probability distribution function \citep[PDF;][]{kim2003, padoan2011, hennebelle2011, federrath2013} which is parametrised via the Mach number $\mathcal{M}$. The self-gravity -- that becomes important at high-densities -- creates a power-law tail in the density PDF, with slope $\alpha=-1.54$ for the volume weighted PDF \citep{kainulainen2009,girichidis2014, stutz2015,schneider2016}. In the following we assume a fiducial Mach number $\mathcal{M}=10$ \citep{pallottini2019, leung2019}, and we indicate the volume-weighted density PDF with $P_V$. Here we summarize the key points of our calculation, while we refer the interested reader to \citet{vallini2018} for a detailed explanation of the algorithm adopted to compute CO emission from each GMC.

In each GMC, we assume that clumps in the high-density tail have size given by the Jeans length $r_i = r_J(n,T)$, while those in the lognormal part of the PDF have length scale related to the volume-weighted density PDF by  $r_i=(\int_{n_i-\delta} ^{n_i+\delta}P_V(n )dn)^{1/3}$.
The volume filled by structures with density $n_i$ is $V_i=(4/3) \pi r_i^3$. The differential number of clumps of a GMC with total volume $V_{tot}$ can be written as
\begin{equation}
d N_{cl} = (V_{tot} /V_i) {\rm d}P_V \,,
\end{equation}
thus the total CO luminosity emission of the $J\rightarrow J -1 $ transition from the GMC is:
\begin{equation}
\label{cotot}
L^{tot}_{\rm CO,J} = \int F_{\rm CO, J}(n_i, N_i, f_{rad})\times 4 \pi r_i^2 dN_{cl}.
\end{equation}

In the previous equation $F_{\rm CO, J}(n_i, N_i, f_{rad})$ is the CO flux computed with the version c17.00 of {\small CLOUDY} \citep{ferland2017} as a function of the gas element (i) density, $n_i$, (ii) column density, $N_i=n_i r_i$, and (iii) the impinging radiation flux ($f_{rad}$). The latter is produced by star-formation -- and parametrised in term of the FUV impinging flux ($G_0$, see eq. \ref{eq:g0}) - and, if the AGN is present, this contribution is added and parametrised in term of the X-ray flux in the 1-100 keV band ($F_{X}$, see eq. \ref{eq:fx}).

The previous equation represents an updated version of the \citet{vallini2018} model that in its original form included the FUV illumination only.
In the next section we outline the details of the PDR and XDR radiative transfer calculations.

\subsection{Radiative transfer}\label{sec:rt_inside_clouds}

\subsubsection{PDR models}\label{sec:PDR_modelling}
To compute the $F_{\rm CO, J}(n, N_i, G_0)$ we run a total of $130= (13 \times 10)$ PDRs models assuming a 1-D gas slab of constant density $n$, with density in the range $\log n=[0, 6]$ (13 models, 0.5 dex spacing) illuminated by stellar radiation only. The spectral energy distribution (SED) of the impinging radiation is obtained with the stellar population synthesis code Starburst99 \citep{leitherer1999} which we scaled to obtain Habing fluxes at the gas slab surface in the range $\log G_0=[0, 5]$ (10 models 0.5 dex spacing).
We assume a Salpeter Initial Mass Function in the range $1-100\,\rm M_\odot$, and we consider a continuous star formation mode with an age of the stellar population of 10 Myr. We adopt the Geneva standard evolutionary tracks \citep{schaller1992} with metallicity $Z_\star = \,\zsun$, and Lejeune-Schmutz stellar atmospheres which incorporate plane-parallel atmospheres and stars with strong winds \citep{lejeune1997, schmutz1992}. We also set the gas metallicity to $Z=Z_{\odot}$.

For the gas slab we assume the standard ISM metal abundances (\texttt{abundances ISM}) stored in Cloudy. We adopt dust ISM grains (\texttt{grains ISM}) with a size distribution and abundance appropriate for the ISM of our galaxy \citep{mathis1977}; this includes both a graphite and silicate component. 

Our model includes also the cosmic microwave background (CMB) radiation at $z=6$. The far-infrared line luminosity from high-$z$ galaxies is in fact highly affected by the increased temperature of the CMB which, one the one hand, represents a stronger background with respect to that at $z=0$, thus lowering the observed flux of the line\citep[e.g][]{obreschkow2009, dacunha2013, vallini2015,pallottini2017}, on the other hand, it keeps the GMCs warmer\citep{vallini2018}.
Finally, we adopt the default {\small CLOUDY} prescriptions for the CR ionization rate background ($\zeta_{\rm CR}$; CRIR, hereafter) $\log(\zeta_{\rm CR}/{\rm s^{-1}})\approx -15.7$ \citep{indriolo2007}. The $\rm H_2$ secondary ionisation rate is $=4.6\times10^{-16}\rm s^{-1}$ \citep{glassgold1974}. As the CRIR is a fundamental parameter that may have strong effects on the gas temperature and chemical composition at high densities, we discuss the impact of our assumption on the CO luminosity inferred with our model in the Sec \ref{sec:RT_CRs}.

\subsubsection{XDR models}\label{sec:XDR_modelling}

We adopt the same approach outlined in the previous section, i.e. we compute $F_{\rm CO, J}(n, N_i, F_X)$ assuming a 1-D gas slab of constant density $n$, $\log n=[0, 6]$ (13 models, 0.5 dex spacing) and all the other relevant parameters ($Z$, metal abundances, dust grains) as discussed in Sec. \ref{sec:PDR_modelling}. The difference is the shape of the impinging SED in the XDR models with respect to the PDR ones. In the case of XDR, we choose the ``table XDR" SED stored in Cloudy. This is meant to generate the X-ray SED described by \citet{maloney1996}, i.e. of the form:
\begin{equation}
\label{eq:sed_xray}
    f_\nu \propto \left( \frac{E}{100\, \rm{keV}} \right)^{-0.7}
\end{equation}
over the energy range $1-100$ keV. The radiation field has negligible intensity outside this range. It is implicitly assumed that the FUV radiation is dominated by star formation distributed over the disk of the galaxy. 
Any other component of the AGN radiation, e.g. the FUV radiation, is completely absorbed by the intervening material due to the torus and to the shadowing of interveining GMCs.

The incident SED is normalized so that the $1-100\, \rm keV$ X-ray flux at the cloud surface is in the range $\log (F_X/{\rm erg\,s^{-1}}) = [-2.0, 2.0]$ (9 models, 0.5 dex spacing), to fully cover the range generally considered in standard XDR modelling such as those presented by \citet{maloney1996, meijerink2007}.

\begin{figure*}
\includegraphics[scale=0.49]{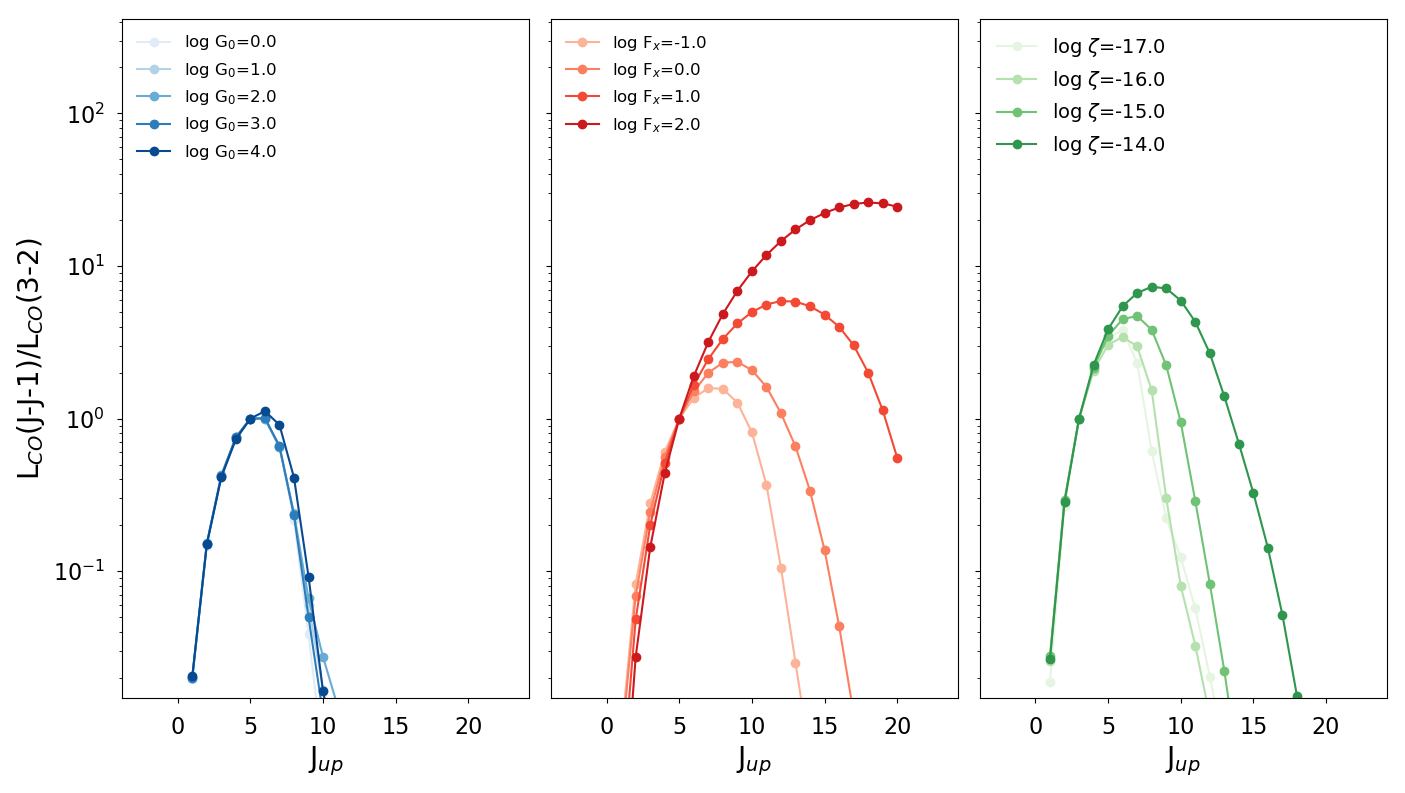}
\caption{The CO SLED luminosities normalized to the CO(3--2) transition of our fiducial GMC. The CO SLED is obtained for the GMC illuminated by stellar radiation (left) with FUV flux in the range $\log G_0=[0, 4]$, by AGN radiation (center) with an X-ray flux at the GMC surface in the range $\log F_{x}=[-1.0, 2.0]$, or assuming an increasing CRIR (right) in the range $\log \zeta_{CR}=[-17, -14]$. \label{fig:COSLED_singlecloud}}
\end{figure*}

\subsubsection{Models with variable cosmic-ray ionization rate}
\label{sec:RT_CRs}
Our fiducial PDR and XDR simulations assume the galactic cosmic ray ionization rate per hydrogen atom derived by \citet{indriolo2007} analyzing se\-ve\-ral lines of sight towards diffuse interstellar clouds in the Milky Way ($\log(\zeta_{\rm CR}/{\rm s^{-1}})\approx -15.7$). Note however that the determination of the cosmic ray ionization rate (CRIR) both in the Milky Way and in external sources, is still highly debated. For instance, in our Galaxy, substantial differences are found between the CRIR measured towards diffuse molecular gas, and that inferred for dense molecular gas \citep[e.g.][]{neufeld2017}, with the former being typically an order of magnitude larger than the latter. Moving towards external galaxies, there is evidence that starburst galaxies such as M82, and ultraluminous infrared galaxies are powerful particle accelerators able to achieve CR energy densities orders of magnitude higher than what is measured in the Milky Way. This results in a cosmic ray ionization rate per atom up to several dex larger than the standard galactic value \citep[e.g.][]{voelk1989, papadopoulos2010, persic2010}
As pointed out by \citet{persic2010} the CR production rate can be approximated being directly proportional to the recent SFR.

The actual CRIR in high-$z$ galaxies is essentially unknown, owing to e.g. the unconstrained magnetic field in the ISM of these very distant objects. Nevertheless, given the starburst nature of high-$z$ sources, it is reasonable to expect an increased CRIR with respect that observed in the Milky Way. An increase of the CRIR always boosts the gas temperature, because the heating provided by CRs is proportional to the cosmic rate ionisation rate. At low densities ($n\approx10^2\rm \, cm^{-3}$), a higher CRIR also causes a decrease of the CO abundance. This is the result of the the dissociative charge transfer reactions of the CO molecules with He$^+$ ions. The ionised carbon produced by these reactions is then converted into neutral carbon (CI). The net effect is that the CO emissivities decrease, because the larger gas temperatures are not enough to compensate for the CO abundance drop.
At much higher densities ($n
\approx 10^5\rm \, cm^{-3}$) the situation is rather different, because the CO abundance drop is smaller and it is compensated by the larger temperatures. As pointed out by e.g. \citet{bisbas2015}, this ultimately boosts the CO emissivity \citep[see also the Appendix of][]{vallini2018}.

To test the impact of the CRIR variation on the CO line emissivity and on the shape of the CO SLED, we run a set of PDR models with varying CRIR in the range $\log(\zeta_{\rm CR}/{\rm s^{-1}}) =[-17.0, -14.0]$ (7 models, 0.5 dex spacing). As for the standard PDR models, we assume a 1-D gas slab of constant density $n$, $\log n=[0, 6]$ (13 models, 0.5 dex spacing), illuminated by stellar radiation only. The spectral energy distribution (SED) of the impinging radiation is the same of the standard PDR models (see Sec. \ref{sec:PDR_modelling}) scaled to have an Habing flux $G_0=100$ at the gas slab surface. In what follows, we will refer to models with varying cosmic rate ionization rate described in this section as \emph{CRIR models}.

\subsection{Coupling RT-models with the GMC structure}
\label{subsec:complete_model}
For all the runs described in the previous Sections,
{\small CLOUDY} computes the radiative transfer through the slab up to a hydrogen column density $N_{\rm H}=10^{23}\,{\rm cm^{-2}}$. 
This stopping criterion is chosen to fully sample the molecular part of the illuminated slab, typically located at $N_{\rm H} > 2\times10^{22}\,\rm{cm^{-2}}$ \citep{mckee2007}. 
The output of each (PDR, XDR, or CRIR) run is the CO line emissivity ($\varepsilon_{\rm CO, J}$) of the various $J \rightarrow J - 1$ transitions up to $J=20$:
\begin{equation}\label{COemissivity}
F_{\rm CO, J} = F_{\rm CO, J}(n_i, N_i, G_0 {\rm \,\, or \,\, } F_X {\rm \,\, or\,\, }\zeta_{C}).
\end{equation}
We plug this in (eq. \ref{cotot}) that returns the total CO luminosity ($L_{\rm CO}$) of the cloud. 

In Figure \ref{fig:COSLED_singlecloud} we plot the CO SLED normalized to the CO(3--2) transition of our fiducial GMC, when the cloud is illuminated by stellar radiation, by AGN radiation, or increasing the CRIR. We note that, as expected for PDR models, the CO SLED peaks around the $J=6$ transition and the shape is almost independent on the $\log G_0$ flux, as obtained in \citet{vallini2018}. For XDR models, instead, the CO SLED peak shifts from $J=7$ for extremely low X-ray impinging fluxes ($\log (F_{X}/{\rm erg s^{-1}})=-1.0$), to $J>20$ in the extreme case of an impinging X-ray flux $\log (F_{X}/{\rm erg s^{-1}})=2.0$. This behavior is in agreement with expectations from previous XDR models \citep[e.g.][]{meijerink2007}. Finally, the signature of an enhanced CRIR begins to show up as a shift of the CO SLED ($J>10$), only in the case of $\log \zeta_{CR}=-14.0$ i.e. 30x greater than the fiducial one \citet{indriolo2007}.

\section{Testing the model in the local Universe}\label{sec:lowzvalidation}
\begin{figure*}
\includegraphics[scale=0.64]{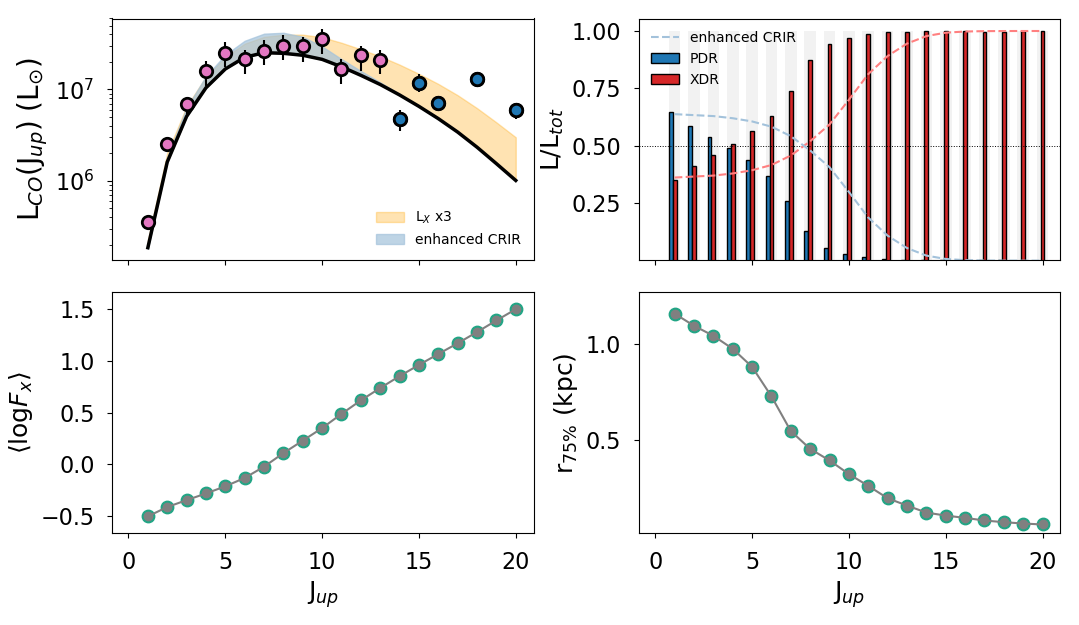}
\caption{\textit{Upper left panel:} CO line observations in Mrk 231. The CO rotational transitions from $J=4$ to $J= 13$ (pink points) are taken from Herschel SPIRE-FTS data presented in \citet{vanderwerf2010}, while CO $J>13$ data (blue points) are obtained from PACS by \citet{mashian2015}. The black dashed line is the CO SLED obtained by summing the contribution from our fiducial GMCs distributed according to the observed gas density profile in Mrk 231 \citet{downes1998}. The light blue shaded region represents the resulting CO SLED obtained by substituting the PDR models with enhanced CRIR models. The orange shaded region represents instead the CO SLED obtained by increasing by a factor of 3 the intrinsic X-ray luminosity of the AGN. \textit{Upper right panel}: Relative contribution to the various CO rotational transitions of 'PDR' models ($G_0=2.5$, blue bars), and 'XDR' models (red bars). With thin dashed lines we show the variation of the relative contribution in the case of PDR with enhanced CRIR (light blue), and XDR (light red).
\textit{Lower left panel:} The luminosity weighted average of the X-ray flux $\langle F_x \rangle$
contributing to each CO line emission.
\textit{Lower right panel:} Galactocentric radius at which the luminosity of a given CO transition reaches 75\% of the total value.} 
\label{fig:COSLED_Mrk231}
\end{figure*}
The model including the effect of FUV photons only, has been already validated in low-$z$ galaxies as discussed \citet{vallini2018}, thus here we proceed with a further sanity check to test whether the inclusion of the XDR component allows us to reproduce the CO SLEDs of well studied AGN host galaxies.

As a benchmark case, we consider the nearby Seyfert galaxy Mrk 231 for which many authors studied in detail the CO excitation mechanism \citep[e.g.][]{papadopoulos2007, vanderwerf2010, mashian2015}.\footnote{As a caveat it is necessary to point out that Mrk 231 shows also a powerful AGN-driven outflow whose impact on the spectral shape of the CO lines has been studied in details by \citet{cicone2012} and \citet{feruglio2015}.} 
The first analysis of the CO SLED excitation in Mrk 231 was limited to observations of the CO ladder up to the CO(6-5) \citep{papadopoulos2007}. \citet{papadopoulos2007} was able to reproduce the Mrk 231 CO SLED exploiting a two component large velocity gradient (LVG) model with different densites ($n\approx10^3 \rm \, cm^{-3}$ and $n\approx10^4 \rm \, cm^{-3}$). Subsequently \citet{vanderwerf2010}, by adding the SPIRE-FTS data up to $J=13$, found that an additional XDR component was needed in order to fit the excitation of high-J CO lines. The best fit included two PDRs -- accounting for the star forming molecular disk (${\rm log} \,n=3.5$) with embedded clumps of dense molecular gas (${\rm log} \,n=5.0$) exposed to strong FUV radiation field (${\rm log}\, G_0=2.5-3.5$) -- plus an XDR accounting for the emission of CO $J>8$ transitions excited in the dense (${\rm log} \,n=4.0$) molecular disk of $\approx 200$ pc  illuminated by X-ray photons (${\rm log} \,F_X\approx 1.5$) from the AGN. The mass associated to the PDR and XDR are $M_{\rm H_2} \approx 6 \times 10^9 \, \rm M_{\odot}$ and $M_{\rm H_2} \approx 2 \times 10^9 \, \rm M_{\odot}$, respectively. More recently, by including also $J>13$ CO observations obtained with PACS, \citet{mashian2015} found that CO lines up $J=11$ in Mrk 231 could be produced by a two component LVG model composed by a cool ($T=50\,\rm K$, $n=10^{3.8}\, \rm{cm^{-3}}$, $M=10^9 \rm M_{\odot}$) and a warmer (denser) molecular phase ($T=316\,\rm K$,  $n=10^{4.2}\, \rm{cm^{-3}}$, $M=10^{8.4} \rm M_{\odot}$). The two component LVG model instead failed to reproduce CO $J>11$.

We now exploit our model to test its ability in reproducing the CO SLED of Mrk 231, and in providing insights on the heating mechanisms driving the CO emission in the galaxy. As pointed out in Sec. \ref{sec:internal_structure}, our model features a physically motivated internal density structure of GMCs, therefore we do not need to use different components at different densities when computing the CO line luminosity. Moreover, the XDR and/or PDR behavior of each cloud depends solely on its actual location with respect to the central AGN. The physics and chemistry of GMCs located close to the galactic center is likely to be more influenced by the X-rays, while GMCs located at greater distances from the central BH, are expected to be more strongly affected by the FUV photon flux that we assume throughout the galaxy. 
To compute the expected CO SLED with our model we proceed as follows:
\begin{itemize}
\setlength\itemsep{1em}
\item[1.] We distribute the GMCs over a disk with a radial mass profile that reproduces that observed in Mrk 231 by \citet{downes1998}: $M_{\rm H_2}= (1.8,\, 3.1,\, 4.0) \times 10^9 \, \rm M_{\odot}$ at $r=460, 850, 1700$ pc, and in good agreement with more recents total molecular mass derivation for Mrk 231, $M_{\rm H_2}\approx 5 \times 10^9 \, \rm{M_{\odot}}$ \citep{vanderwerf2010, cicone2012}. The height of the disk is set to $H=50\rm \, \rm pc$ \citep{teng2014}.

\item[2.] We set the 1-100 keV intrinsic X-ray luminosity of the AGN to $L_X=1.5 \times 10 ^{43} \,\rm erg\, s^{-1}$ extrapolating (see eq. \ref{eq:sed_xray}) the most recent measurement in the range $0.5-30$ keV \citep[$L_{X}(0.5-30\rm keV)\approx 1 \times 10^{43}\rm \, erg\, s^{-1}$,][]{teng2014}. The obscuring torus \citep[$N_H=1 \times 10^{23}\rm \,cm^{-2}$][]{teng2014} is assumed to be tilted with respect to the molecular disk of about $\Omega=55^{ \circ}$ inclination \citep{teng2014,feruglio2015}. This translates into a $1-100$ keV X-ray flux at $\approx 160\rm \, pc$ from the galaxy center $F_X\approx 1^{+0.8}_{-0.6}\rm \, erg\, s^{-1}\, cm^{-2}$ due to the variation of viewing angle of the central AGN. The $F_X$ we get is $\approx 10$ times lower than the X-ray flux adopted by \citet{vanderwerf2010} for their XDR component of $\approx 160$ pc size ($F_X\approx 28 \rm \, erg\, s^{-1} \, cm^{-2}$) which fit the high-J CO line emission of the observed CO SLED.
Note however that \citet{vanderwerf2010} considered a higher intrinsic $L_{X}(2-10\, {\rm keV})=6 \times 10^{43} \rm erg \, s^{-1}$ luminosity for Mrk 231 \citep[i.e. that derived by][]{braito2004}, and pointed out that $F_X\approx 28\rm \, erg\, s^{-1} \, cm^{-2}$ to a distance of $\approx 160\rm \, pc$ from the nucleus could be produced by this AGN only \emph{ignoring absorption}.

\item[3.] Considering that the SFR measured in the circumnuclear starburst of Mrk 231 is $\approx 200\rm \, M_{\odot}\, yr^{-1}$ \citep[e.g.][]{taylor1999, cicone2012}, by adopting eq. \ref{eq:g0}, we derive a FUV flux ${\rm log}\, G_0\approx 2.5$. We set this as fiducial impinging `stellar' radiation for all the GMCs in our model. Note that this value is in perfect agreement with the $G_0$ inferred by \citet{vanderwerf2010} by fitting low-$J$ transitions of the CO SLED with their 'diffuse PDR' component. 

\item[4.] For each GMC we compute the resulting CO SLED (see Fig. \ref{fig:COSLED_singlecloud}) according to the impinging X-ray radiation given its distance from the central AGN, and FUV radiation as discussed in the previous point.

\item[5.] The total emission for all the CO rotational lines in the range $J_{up}=1-20$, is obtained by summing the CO luminosity of each GMC.
\end{itemize}

The result of the procedure is depicted in the upper left panel of Fig. \ref{fig:COSLED_Mrk231}. There, the points represent the observed CO luminosities up to the CO(20--19) rotational transition in Mrk 231, and the black line indicates the CO SLED obtained with our fiducial model. We note that the output from our calculation is in very good agreement the with the observations up to J=16, even though we fail in reproducing the two uppermost transitions CO(18--17) and CO(20--19) even if we increase by a factor of 3 the intrinsic X-ray luminosity of the central AGN. The starburst composite nature of Mrk231 likely implies larger CR ionization rates with respect to the Milky Way \citep{indriolo2007}. We have therefore included in Fig. \ref{fig:COSLED_Mrk231} the CO SLED obtained by substituting the PDR models described above, with those having enhanced CRIR ($\log(\zeta)=-14.5$). The result of this procedure is highlighted with a light blue shaded area. As expected from Fig. \ref{fig:COSLED_singlecloud}, the enhanced CRIR models boost CO lines ($J\leq 9$), while leaving the very high-J CO lines unaffected.

We further analyze the CO SLED, exploiting the power of our model in understanding both the dominant heating mechanism (XDR vs PDR), and the prevalent X-ray flux and location of the GMCs contributing to a given transition. More precisely, in the upper right panel of Fig. \ref{fig:COSLED_Mrk231}, we compute the contribution of XDR vs PDR component. We recall that each cloud in our model is illuminated by a constant $\log G_0=2.5$, and by an X-ray flux $F_X$ determined by the location of the GMC within the disk.
From Fig. \ref{fig:COSLED_Mrk231} we note that CO transitions with $J\geq 9$ are totally dominated by the XDR component. CO transitions with $5<J<11$ are mostly produced from XDR, with a residual contribution of PDR. More precisely, the PDR fraction for transitions with $J>6$ is always below 0.25, and drops to zero already at $J>8$. The PDR component instead, accounts for $\approx 60\%$ of the CO(1-0) and CO(2-1) emission, with the rest provided by the XDR component.
In the case of PDR with enhanced CRIR (dashed lines Fig. \ref{fig:COSLED_Mrk231}) the XDR completely dominates the CO emission for $J\geq13$. The enhanced CRIR models account for $>50\%$ of the CO fluxes for transitions with $J<8$. The CO(1-0) and CO(2-1) emission are not affected by the enhancement of the CRIR and the relative fraction of PDR vs XDR remains substantially unchanged.

If we focus our attention on the spatial location of the emission (see lower right panel of Fig. \ref{fig:COSLED_Mrk231}), we note that the galactocentric radius ($r_{75\%}$) that contains 75\% of the total emission of the various CO transitions decreases from $r_{75\%}\approx 1.25 \, \rm {kpc}$ for the CO(1--0) and CO(2--1) up to $r_{75\%}< 0.5 \, \rm {kpc}$ for $J>10$. The closer location to the galaxy center of GMCs contributing to high-J transitions, translates into correspondingly higher X-ray fluxes contributing to the overall emission of each CO transition. This is quantified in the lower left panel of Figure \ref{fig:COSLED_Mrk231} where we show the luminosity weighted average of the X-ray flux $\langle F_x \rangle$ contributing to the various rotational transitions. 
The spatial location of the CO emission as obtained from the enhanced CRIR models produces a more extended CO emission for the 3<J<9, with a maximum increase of a factor $\approx 1.5$ for $J=6$.

To conclude, our analysis is thus in agreement with previous studies. We naturally find without any a priori assumption -- excluding that concerning gas mass distribution -- that: (i) PDRs dominate the emission of the lower $J$ lines, and the emission arises from larger distances from the center of the galaxy, (ii) the $J>10$ levels are due to GMC located in the central part of the galaxy where the chemistry and physics of molecular emission is highly affected by the X-ray flux.
After this encouraging test, in the next section we will define our fiducial model for a typical high-$z$ galaxy, and we will then discuss the effect of various parameters on the expected CO SLED.

\section{Application in the EoR}\label{sec:parameter_effect}
\begin{figure*}
\includegraphics[scale=0.55]{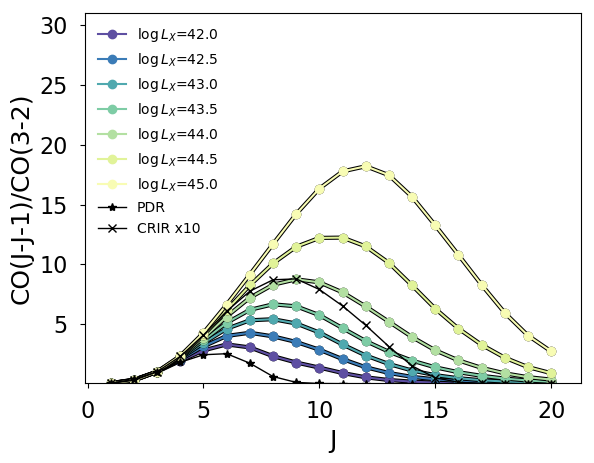}
\includegraphics[scale=0.55]{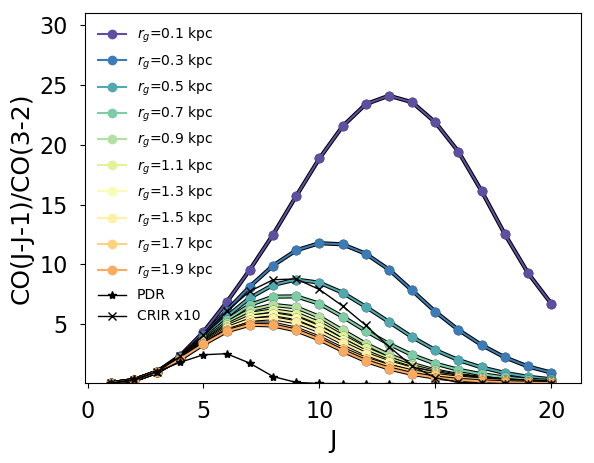}
\includegraphics[scale=0.55]{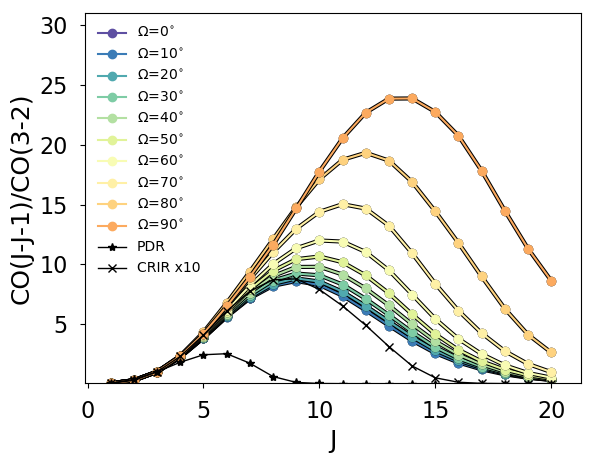}
\includegraphics[scale=0.55]{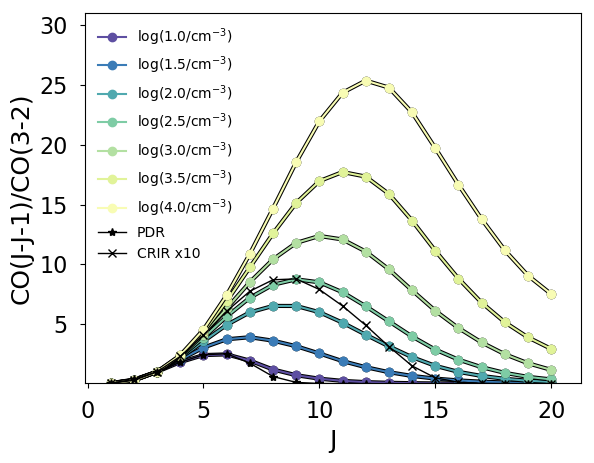}

\caption{\textit{Top-left:} The CO SLED luminosities normalized to the CO(3--2) line, for the fiducial disk model (see Tab. \ref{tab:fid_model}) when varying the X-ray luminosity of the central AGN. \textit{Top-right:} Impact on the CO SLED of different sizes of the molecular disk $r_g$, keeping the X-ray luminosity fixed to the fiducial value, $L_{X}=10^{44} \rm{erg \, s^{-1}}$. \textit{Bottom-left:} Impact of different inclination angles of the obscuring torus for $L_X$ fiducial. \textit{Bottom-right:} Impact of different GMC mean densities keeping all the other parameters fixed to the fiducial ones. In all the plots, the black lines with stars and crosses represent the CO SLED resulting for the fiducial PDR calculation, and that resulting from the CRIR model calculation considering a cosmic ray ionization rate 10 times greater than the fiducial one.\label{fig:impact_parameters}}
\end{figure*}
As pointed out in the Introduction, high redshift galaxies are known to be more compact \citep[e.g][]{tacchella2016, bouwens2017, jiang2013, oesch2010} than lower redshift sources. They have also disturbed morphology \citep[e.g][]{carniani2018} owing to the high merging rate that characterizes galaxies at early epochs. Moreover, numerical simulations -- as well as recent ALMA observations -- show the presence of compact rotating disks \citep[e.g.][]{jones2017, smit2018}, which are often perturbed by the accretion of the merging clumps and satellites \citep{kohandel2019}.

We describe a high-$z$ galaxy considering two extremes: 1) a disk model so that the total gas mass 
($M_{\rm H_2}=1.7\times10^8$), the size of the gas disk ($r_g=504$ pc), the mean density of GMCs ($\log (n_0)=2.5$), and the height of the disk ($H=194$ pc) are in agreement with those found in state-of-art zoom-in cosmological simulation presented in \citet{pallottini2017b}. The case of the fiducial perturbed morphology model, is instead obtained considering a spherical model, with clouds distributed over a spherical volume up to $r_g=504$ pc. The values are reported in Tab.\ref{tab:fid_model}.
In what follows we want to asses if -- and to what extent -- the possible presence of a (low-luminosity) AGN in the center of such a galaxy would affect the CO emission, and the line ratios. To do that we have to assume a fiducial $L_X$ luminosity for the AGN. The procedure adopted is the following:
\begin{itemize}
    \setlength\itemsep{1em}
    \item Starting from the stellar mass of the fiducial galaxy, $M_\star=2.6 \times 10^{10} \, \rm M_{\odot}$ \citep{pallottini2017b}, we set the BH mass to $M_{\rm BH}\approx 2.6 \times 10^7 \, \rm M_{\odot}$, assuming the typical $M_{\rm BH}/M_\star \approx 10^{-3}$ ratio observed in the local Universe \citep[e.g.][]{marconi2003, haring2004, reines2015}. As a caveat, we note that whether this ratio increases ($>10\times$) with redshift is still a matter of debate both observationally \citep[e.g.][]{decarli2018, izumi2019}, and theoretically \citep[e.g.][]{barai2018, lupi2019}.
   \item From $M_{\rm BH}$, assuming Eddington accretion, we compute the bolometric luminosity $ \left(\frac{L_{bol}}{{\rm erg\, s^{-1}}}\right) = 1.26 \times 10^{38} \left(\frac{M_{\rm BH}}{M_{\odot}}\right)
        \approx 3 \times 10^{45}\left (\frac{M_{\star}}{2.6 \times 10^{10} M_{\odot}}\right)$.
 \item By adopting the 2-10 keV $L_X-L_{bol}$ bolometric correction from \citet[][see their eq. 2]{hopkins2007}, we infer the fiducial X-ray luminosity $L_X\approx 10^{44}\rm \, erg \, s^{-1}$. 
\end{itemize} 

In what follows, we will isolate, one by one, the effects of: 1) the increasing X-ray luminosity, 2) the compactness of the galaxy, 3) the relative angle between the torus and the galactic plane 4) the variation of the mean density of GMCs and, 5) the geometry
of the molecular cloud distribution within the galaxy.

\subsection{X-ray luminosity from the central engine}
Let us assume the fiducial disk model model (see Table 1) and compute the resulting CO SLED normalized to the CO(3--2) transition, as shown in the top left panel of Fig. \ref{fig:impact_parameters}. The fiducial model assumes the worse case scenario in which the tilt angle $\Omega = 0^{\circ}$. In this case the maximum obscuration of the torus is in the direction of the molecular disk.
We can notice that, as expected, with increasing X-ray luminosity from the central AGN, the CO SLED is more and more excited. For comparison we plot the resulting CO SLED that one would obtain with the PDR models only, for the same fiducial distribution, and that resulting from CRIR models with 100x higher cosmic ray ionization rate. From the figure appears clear that if the central luminosity of the AGN is $\log(L_{\rm X}/{\rm erg\, s^{-1}})> 44$ one can recover the presence of the AGN from the CO SLED shape without the risk of other sources of excitation mimicking the same behavior. For $\log (L_{\rm X}/{\rm erg\, s^{-1}})< 44$ , instead, the CO SLED is comparable to that obtained for the enhanced CRIR calculation. The CO SLED obtained for the PDR only calculation is always less excited than all the other models CRIR and/or XDR models.

\subsection{Molecular distribution size}
The size of the molecular distribution $r_g$ has obviously an impact on the CO SLED because if the galaxy is compact the molecular gas is, on average, located close to the central AGN and thus the effect of the X-ray is enhanced. In this test, we keep the X-ray luminosity at our fiducial value of $L_X=10^{44} \rm \, erg\, s^{-1}$. The top-right panel of Fig. \ref{fig:impact_parameters} shows that, that if  $r_g$ increases from $0.1$ kpc, up to $\approx 2$ kpc, the excitation of the CO SLED decreases continuously. We note that already for $r_g=500-700\rm \, pc$ the effect of X-rays is practically indistinguishable from that of an enhanced CRIR. This point is critical: given that high-$z$ galaxies are much more compact that low-$z$ ones, this may increase the chance of getting the signature of intermediate luminosity AGN through the CO SLEDs. The same AGN, in low-redshift galaxies with much larger molecular disks, would have had less important effect on the molecular gas.

\subsection{Tilt angle}
The tilt of the torus with respect to the molecular disk, and thus with respect to the location of GMCs in the galaxy, is a very important parameter shaping the CO SLED excitation. This is shown in the bottom-left panel of Fig. \ref{fig:impact_parameters} in which we plot the CO SLED from the fiducial model, assuming an X-ray luminosity
of the AGN  $L_{X}=10^{44} \rm{erg \, s^{-1}}$. As can be appreciated from the bottom left panel in the Figure \ref{fig:impact_parameters}, for $\Omega\leq 50^{\circ}$ the CO SLED is only slighlty affected by the rise of the inclination angle. In this case, as for the fiducial model with $\Omega=0^{\circ}$, the CO SLED is almost indistinguishable from that obtained with an increased CRIR. Instead, if $\Omega\geq60^{\circ}$, the CO SLED excitation increases very steeply. Let now focus our attention on the CO(6--5) transition, which is the first CO line accessible from $z>6$ with ALMA. The excitation of this line, even for the extreme case of $\Omega=90^{\circ}$ inclination, is always almost indistinguishable from that obtained for the enhanced CRIR model. Instead, our model suggests that pinpointing lines with $J\geq 10$ is more promising in term of disentangling the effect of the AGN, as a tilt inclination of $\Omega\approx 70^{\circ}$ results in a factor $\approx 2$ more excited line than the corresponding line obtained for the enhanced CRIR model. 

\subsection{Impact of GMC mean density}

Increasing the mean density of the GMCs results in a boost and shift of the peak of the CO SLED towards higher-J transition. This is in line with expectations, because high-J CO rotational transitions trace increasingly high densities \citep[e.g.][]{weiss2007}. We recall here that the GMCs of our model feature an internal density field. At fixed Mach number, an increase in the mean density, $n_0$, results into a translation towards higher densities of the PDF describing their internal density distribution. As a consequence, the CO SLED gets more excited.
By looking at the bottom right panel in Fig. \ref{fig:impact_parameters}, it is clear that the impact of density variations is comparable to that of X-ray luminosity variations. Indeed, an increase of 1 dex from the fiducial density (from $n_0=10^{2.5}\, \rm cm^{-3}$ to $n_0=10^{3.5}\, \rm cm^{-3}$) produces the same effect on the CO SLED shape of an increase of 1 dex of the X-ray luminosity (from $L_X=10^{44}\, \rm erg\, s^{-2}$ to $L_X=10^{45}\, \rm erg\, s^{-2}$). Note that densities of the order of  $n=10^{3.5}-10^{4.5}\rm \, cm^{-3}$ are typical in sources with very excited CO SLEDs such as high-$z$ QSO which are also expected to have high BH masses and thus correspondingly high X-ray fluxes \citep[e.g.][]{venemans2017molecular, novak2019, wang2019}.

\subsection{Geometry}
\begin{figure}
    \centering
    \includegraphics[scale=0.55]{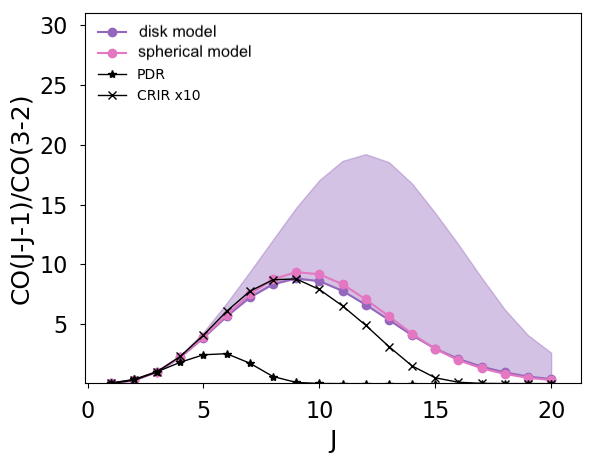}
    \caption{The CO SLED luminosities normalized to the CO(3--2) resulting for a spherical distribution (spherical model), or for the disk distribution (disk model). For the disk model, the shaded area highlights the possible variation of the CO SLED from $\Omega=0^{\circ}$ to $\Omega=70^{\circ}$ deg. Due to the isotropy, the spherical model is not affected by the orientation of the obscuring torus.}
    \label{fig:geometry}
\end{figure}

In Fig. \ref{fig:geometry} we investigate the impact of the distribution of the clouds in the galaxy. More precisely, we plot the resulting CO SLED for both the disk and spherical models. We note that the isotropy of the spherical model makes the orientation of the obscuring torus irrelevant in this configuration.
In the worst-case scenario in which the torus is in the plane of the galactic disk (i.e. $\Omega=0^{\circ}$) then the CO SLED resulting from the spherical distribution and that obtained for the disk distribution are almost indistinguishable. If instead the inclination angle of the torus with respect to the galactic plane increases, then the CO emission from the disk model model is increasing and always exceeding that obtained for the spherical distribution. 

\begin{table}
 	\centering
 	\caption{Parameters of the high-$z$ galaxy fiducial model, for spherical and disk model cases. From left to right: $M_{\rm H_2}$ in $M_{\odot}$, radius of the molecular disk $r_g$ in pc, height of the molecular disk $H$ in pc. We report also the fiducial properties of each GMC: mean density (in $\rm cm^{-3}$), mach number $\mathcal{M}$.}
	\label{tab:fid_model}
	\begin{tabular}{l|ccc|cc}
 		\hline
 		&$M_{\rm H_2}$ & $r_g$ & $H$ & $\log n_0$ & $\mathcal{M}$ \\
		\hline
	disk model	&$1.7\times 10^{8}$ & 504 & 194 & 2.5 & 10 \\
    spherical model &$1.7\times 10^{8}$ & 504 & -- & 2.5 & 10 \\
 		\hline
	\end{tabular}
\end{table}

\section{CO/[CII] in high-z galaxies and quasars}
\label{sec:cii_co_QSO}
\begin{figure*}
    \centering
    \includegraphics[scale=0.6]{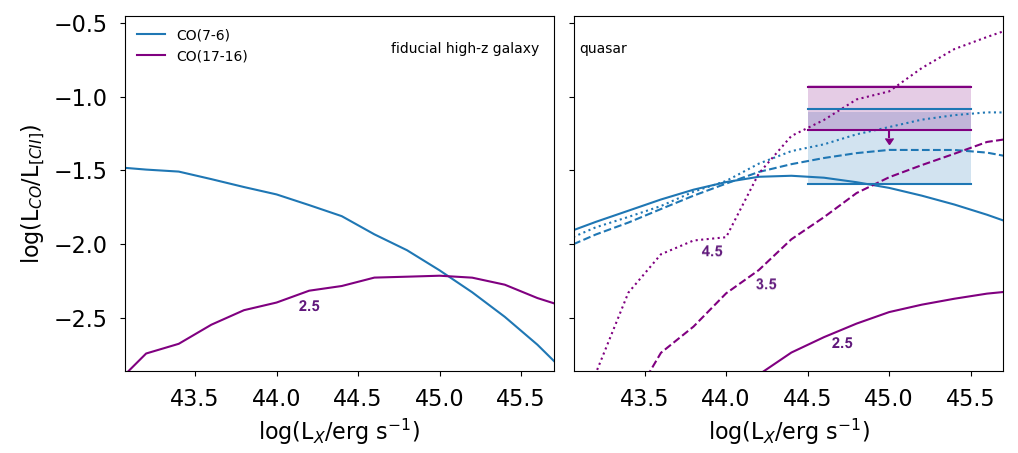}
    \caption{\emph{Left panel:}The CO(7--6)/[CII] (blue line) and CO(17--16)/[CII] (purple line) ratios as a function of the X-ray luminosity for the fiducial high-$z$ model of a normal star forming galaxy (see Tab. 1 for the parameters), assuming $\Omega=0^{\circ}$. In particular we recall that the fiducial GMC mean density is $n_0 = 10^{2.5} \rm \, cm^{-3}$. \emph{Right panel:} Line ratios obtained with our model assuming the properties ($r_g\approx 1.5 \, \rm kpc$, $M_{\rm H2}\approx 10^{10}\rm \, M_{\odot}$) typical of high-$z$ QSO. We test three different mean GMC densities $10^{2.5}\, \rm {cm^{-3}}$ (solid lines) $10^{3.5}\, \rm {cm^{-3}}$ (dashed lines), and $10^{4.5}\, \rm {cm^{-3}}$ (dotted lines). The shaded areas highlights the range of observed ratios (or upper limits) for CO(7--6)/[CII] and CO(17--16)/[CII] in high-$z$ QSOs \citep[J1148+5251, J0305+3150, J0109+3047, J2348+3054, J0100+2802, J2310+1855, J1319+0959, J1342+0928][respectively]{gallerani2014, venemans2017molecular, wang2019, carniani2019:CO, novak2019}. Given that the $2-10$ keV X-ray luminosity ($L_{X(2-10)} \approx 1 \times 10^{45}\rm \, erg \, s^{-1}$) has been measured in J1148+5251 \citep{gallerani2017}, J1342+0928 \citep{banados2018}, and J0100+2802 \citep{nanni2017} only, while the others QSOs are undetected in X-ray, we decide to highlight the X-ray range $L_X=[10^{44.5}-10^{45.5}]\rm \, erg\,s^{-1}$.}
    \label{fig:co_cii_ratios}
\end{figure*}
In the previous Sections we have discussed the impact of geometry, $r_g$, $L_X$, $\Omega$, and $n_0$ on the CO SLED excitation, here instead we focus on the impact of X-ray luminosity on the ratio of two relevant CO rotational transitions (CO(7--6) and CO(17--16)) over the most luminous line in the FIR band: the [CII] at $158 \, \rm \mu m$.  We focus only on these transitions to keep the discussion simple and because (1) the CO(7--6) is the CO line predicted to be the most luminous line in high-$z$ normal star forming galaxies in the absence of AGN \citep{vallini2018}, (2) the CO(17--16) is the nearest CO transition to the [CII] line, thus often falling in the same band of [CII] observations, and can be used to disentangle the presence and influence of AGN.

Predicting the behaviour of the CO/[CII] line ratios is of great importance in the context of high-$z$ galaxy ISM characterization, because with ALMA the [CII] line is routinely detected at $z>6$ \citep[e.g][]{willott2015, capak2015, maiolino2015, knudsen2016,carniani2018b}. It is thus clear that any follow up in CO of high-$z$ galaxies, generally starts with a previously available [CII] detection. The [CII] line luminosity, tracing the outer layers of the GMCs, is computed following the exact same procedure outlined in Sec. \ref{sec:PDR_modelling}, and \ref{sec:XDR_modelling}, i.e. considering 1-D PDR and XDR models and then combining them according the internal density field of the GMC.

In the left panel of Fig. \ref{fig:co_cii_ratios} we plot the CO(7--6)/[CII] and CO(17--16)/[CII] ratios as resulting from our  model of the fiducial high-$z$ normal star forming galaxy, for a fixed tilt angle $\Omega=70^{\circ}$. We consider X-ray luminosities in the range $L_X=[10^{43}, 10^{45.5}]\, \rm erg\, s^{-1}$. All the other parameters have been kept fixed to those quoted for fiducial disk model model in Table \ref{tab:fid_model}. In Fig. \ref{fig:co_cii_ratios} we see that CO(7--6)/[CII] ratio decreases for X-ray luminosity ranging from $L_X=10^{43} \, \rm erg \, s^{-1}$ up to $L_X=10^{45.5} \, \rm erg \, s^{-1}$.  For X-ray luminosities above $L_X=10^{44.5} \, \rm erg \, s^{-1}$ the ratio falls below the CO(17--16)/[CII]. This is due to the the fact that when the X-ray luminosity is high, low-$J$ and mid-$J$ transitions are suppressed because in the low density gas ($\log n\approx 2.5-3.0$) traced by these lines, the CO is dissociated.  Instead, the [CII] line luminosity is boosted by increased temperature in the XDR. This causes a drop in the CO(7--6)/[CII] ratio.
When it comes to high-$J$ lines, such as the CO(17--16) transition, which is excited in the warm dense ($\log n\approx 5-6$) molecular clumps of GMCs, the heating by X-rays up to high (column) densities induces an enhancement in the CO luminosity. However this holds true up to $L_X=10^{44.5}\, \rm erg\, s^{-1}$. Above this limit, we observe the same decreasing trend found for the CO(7--6)/[CII] ratio. Again, this means that the warmer temperature of the gas is not enough to compensate the dissociation of CO molecules.

While high-$z$ normal star forming galaxies remain the main focus of this paper, we further analyze the impact of X-rays on the CO emission for QSO host galaxies, that are more massive ($M_{dyn}\approx10^{11} \rm \, M_{\odot}$, $M_{\rm H2}\approx10^{10} \rm \, M_{\odot}$) and highly star forming (SFR$\approx 1000\,\rm M_{\odot}\, yr^{-1}$), but for which observational data (e.g.  CO(7--6), CO(17--16) and [CII]) are rea\-dily available.

In the right panel of Fig. \ref{fig:co_cii_ratios} we thus report the CO(7--6)/[CII] and CO(17--16)/[CII] obtained with our model distributing the GMCs over a disk with radius $r_g=1.5\, \rm {kpc}$ and up to a total mass of $M_{\rm H2}=1\times 10^{10}\rm M_{\odot}$, which represents the typical [CII] size \citep[e.g.][]{maiolino2007, decarli2018,venemans2017}, and molecular mass observed in $z>6$ QSOs. \citep[e.g.][]{gallerani2014,venemans2017molecular, wang2013, wang2019, novak2019}. In this case, we consider different values for the GMC mean density: $n_0=10^{2.5}\, \, cm^{-3}$ (i.e. the same assumed for the fiducial high-$z$ normal galaxy model), and $n=10^{3.5}, 10^{4.5}\, \rm {cm^{-3}}$ (which are those typically inferred for QSO at high-$z$).
In this situation, the increased gas mass and column densities, allow a larger obscuration of the X-ray photons thus keeping the CO(7--6)/[CII] higher in the $L_X<10^{44.5}\, \rm {erg\,s^{-1}}$ range with respect to the high-$z$ galaxy model at the same density.
The CO(7--6)/[CII] ratio nicely falls through the observed region (shaded blue area) highlighting CO(7--6)/[CII] ratios in the following QSOs: J0100+2802 \citep{wang2019}, J0305+3150, J0109+3047, J2348+3054 \citep{venemans2017molecular}, J1342+0928 \citep{novak2019}. The three different densities do not differ much in term of the predicted CO(7--6)/[CII].
Nevertheless, as already discussed in Sec \ref{sec:parameter_effect}, the increase of the mean density has a strong effect on high-J CO lines and in particular here we see the effect on the CO(17--16). The purple shaded area highlights the \emph{upper limits} on the CO(17--16)/[CII] observed in high-$z$ QSO. The only detection is that in J1148+5251 \citep{gallerani2014}, which represents the upper bound of the purple shaded area, all the other high-$z$ QSOs \citep[J2310+1855, J1319+0959, J1342+0928;][respectively]{carniani2019:CO, novak2019} are \emph{undetected} in CO(17--16) and the upper limits on their CO(17--16)/[CII] ratios fall within the purple shaded area.

\section{Discussion and Conclusions}
\label{sec:conclusions}
In this paper we developed a semi-analytical model able to predict the effect of the presence of X-rays produced by the AGN on the CO SLED of low-$z$ and high-$z$ galaxies. The model accounts for the distribution and internal structure of GMCs around the central AGN. We also incorporated a simplified treatment of the AGN torus obscuration as a function of the inclination angle with respect to the host galaxy disk.
The model has been successfully tested against the well studied low-$z$ Seyfert galaxy Mrk231, for which the AGN impact on the CO SLED has been extensively discussed by \citet{vanderwerf2010}.
More precisely, we were able to fit the observed CO SLED of the Mrk231 by feeding as input parameters the observational constraints on the mass of the gas, the X-ray luminosity of the central AGN, and the gas surface density radial profile. We then applied the model to high-$z$ galaxies which are expected to be more compact than low-$z$ sources. The model allow us to draw the following conclusions:
\begin{itemize}
    \item compact galaxies have enhanced CO SLED excitation with respect to sources with comparable AGN activity but larger disk size.
    \item the effect of enhanced CRIR can be ruled out from the observation of the CO SLED only if the X-ray luminosity of the central AGN exceeds $L_X=10^{44}\rm \, erg\, s^{-1}$, or if the inclination angle between the torus and the molecular disk is $\Omega\geq60^{\circ}$ and one focuses on high-$J$ ($J\geq8$) CO lines. 
    \item the line ratio between high-$J$ CO lines, such as CO(17--16) over the [CII] increases with increasing X-ray luminosity and gas density. For mid-J CO lines - such as the CO(7--6) - the trend is more complex. For X-ray luminosities above $L_X=10^{44.5} \, \rm erg \, s^{-1}$ and mean GMC densities $\log n=2.5$ the mid-$J$ transitions are suppressed because in the intermediate density gas traced by these lines the CO starts to be dissociated. This causes a drop in the CO(7-6)/CII ratio at high X-ray fluxes. If instead the mean GMC density is higher ($\log n = 3.5, 4.5$) the CO(7-6)/CII increases with X-ray flux increasing.
    \item We verified that our model reproduces the observed CO(7--6)/[CII] and CO(17--16)/[CII] ratios and upper limits currently available in high-$z$ QSOs, thus probing the applicability of our approach in the interpretation of future and current observations.
\end{itemize}
Nevertheless, there some caveats to keep in mind that may have impact on our conclusions:
\begin{itemize}

    \item[1.] The impinging FUV flux is assumed to be the same for all the GMCs, and scaled with the fiducial SFR. The impact of this choice is expected to be marginal as varying $G_0$ over 4 order of magnitudes from $\log G_0=0$ to $\log G_0=4$, see Fig. \ref{fig:COSLED_singlecloud}, has small effects on the CO SLED excitation.
    
    \item[2.] Our scheme for the X-ray obscuration produced by the torus is based solely on the inclination angle of a donut-shaped homogeneous absorbing material. This has to be considered as a simplification as both IR and X-ray studies support instead a clumpy structure for the torus \citep[e.g.][]{markowitz2014, ramosalmeida2017}. In particular, \citet{markowitz2014} found X-ray absorption events consistent with eclipses produced by the transit of clumps ($N_H \approx 10^{22-23}\, \rm cm^{-2}$) along the line of sight over timescales of $\approx$days. This affects the X-ray flux -- and thus the $F_X$ parameter entering in the XDR models -- impinging the GMCs in the inner part ($r\leq r_{GMC}$) of the galaxy. There, the absorption ($N_{gas}$) produced by other intervening GMCs -- which is also accounted for in our model (see eq. \ref{eq:absorption_gas}) and dominates at larger distances from the center -- is negligible, and the clumpy nature of the torus should be properly accounted for. 
    We note, however, that the typical timescales for the CO cooling are of the order of $\approx$ kyr \citep[see e.g.][]{krumholz2011}, thus making the overall CO emission not sensitive to variations of $F_X$ over timescales as short as days.
    
    \item[3.] In the present analysis we do not address the possible contribution of shocks to the CO excitation \citep[e.g.][]{hollenbach1989, flower2010, godard2019}. Shock waves in the ISM can in fact produce excitation of the mid-$J$/high-$J$ lines as a consequence of the heating and compression of the molecular gas. The presence of shock induced excitation of CO lines has been invoked for instance to explain the flattening of the CO SLED observed in several local starburst and Seyfert galaxies. \citep[e.g.][]{rosenberg2015}.
\end{itemize}

We conclude commenting on the feasibility of using mid-J and/or high-J lines to reveal the presence of AGN.
As discussed in Sec. \ref{sec:parameter_effect}, a degeneracy exists between increasing $\Omega$, decreasing $r_g$, and increasing $n_0$, that have similar effects on the CO SLED excitation. While $\Omega$ is likely difficult to be derived at high-$z$, independent measures for $n_0$ and $r_g$ can be obtained. The radius of the molecular disk can be constrained with spatially resolved CO observations. While this is certainly a time-expensive task, this has been already proven to be feasible with the (more luminous) [CII] line, that allowed some of the first morphological and dynamical studies of the neutral/molecular gas in normal star-forming galaxies at high-$z$ \citep[e.g.][]{jones2017, smit2018, carniani2018, kohandel2019}. 
The high-density gas can be instead pinned down by targeting dense gas tracers such as the HCN/HCO+ rotational lines \citep{riechers2010} and also exploit their ratio to infer the average X-ray flux impinging the molecular gas \citep[][]{meijerink2007}. Their ratio can be also used to discriminate from the enhanced CRIR mo\-dels. Given that the luminosity of such lines is lower than that of CO transitions, this might be however achievable only for QSOs.

Finally, we can predict the observability of the CO lines in normal galaxies, with and without the AGN. Our model, in the fiducial case of density $\log n_0 = 2.5$, returns CO(7--6)/[CII]$ \approx 1/30$ for the luminosity range typical of low-luminosity AGN ($L_{X}<10^{44}\, \rm erg\, s^{-1}$), while the ratio drops for larger AGN luminosities due to the CO dissociation.
Using the ALMA observing tool, and considering that the typical [CII] luminosity of such a galaxies is $L_{\rm [CII]}\approx 10^{8.5}\rm \, L_{\odot}$ \citep[see the compilation in][]{carniani2018b}, this translates in a predicted observing time of $\approx 20$ hours for the CO(7--6) line. A larger (column) density ($\log n_0 = 4.5$), such as for the case of quasar host galaxies, or induced by the frequent merging events characterizing the assembly of first galaxies \citep[see e.g. the discussion in][]{kohandel2019}, might boost these ratios up to CO(7--6)/[CII]$\approx 1/10$ and CO(17--16)/[CII]$\approx-0.7$, respectively in the high X-ray luminosity in the regime $L_X>10^{45}\, \rm {erg\, s^{-1}}$.
A thorough assessment of the time evolution CO/[CII] ratios as a function of merging events, as well as a complete modelling of the QSO host galaxies will be addressed in a forthcoming paper.

\section*{Acknowledgements}
We thank the anonymous referee for the helpful comments that increased the quality of this paper. LV acknowledges funding from the European Union's Horizon 2020 research and innovation program under the Marie
Sk\l{}odowska-Curie Grant agreement No. 746119.



\bibliographystyle{mnras}
\bibliography{XDR_CO_highz} 


\bsp	
\label{lastpage}
\end{document}